\def\w1jet{$W+1{\rm -jet}$ }

\def\wge1jet{$W+\geq 1{\rm -jet}$ }

\def\wev{$W\to e\nu$ }

\documentclass{ws-procs9x6} 

\begin{document}

\title{W + jet production at CDF}

\author{Andrea Messina}

\address{(on behalf of the CDF collaboration)\\
INFN Sezione di Roma,
Piazzale Aldo Moro, 2, Roma, 00124, Italy\\ 
E-mail: andrea.messina@roma1.infn.it}

\maketitle

\abstracts{The cross section for the inclusive production of W
bosons in association with jets in $p\bar{p}$ collisions at 
$\sqrt{s}=1.96$~${\rm TeV}$ using the Collider Detector at Fermilab (CDF II)
is presented. The
measurement is based on an integrated luminosity of $320$~${\rm pb^{-1}}$, and includes
events with up to $4$ or more jets. In each jet
multiplicity sample the differential and cumulative cross sections with respect to 
the transverse energy of the $i^{th}-$jet are measured. 
For $W+\ge2$ jets the 
differential cross section with respect to the 2-leading jets invariant mass $m_{j_{1}j_{2}}$
and angular separation $\Delta R_{j_{1}j_{2}}$ is also reported.
The data are compared to predictions from Monte Carlo simulations.}

      
The study of jets produced in events containing a $W$ bosons provides 
a useful test of Quantum Chromo-Dynamics (QCD) at high momentum transfers.
Recently a lot of work has been channeled to develop 
sophisticated Monte Carlo programs capable of handling more particle in the final
state at the leading order (LO), or in some cases, next-to-leading order
(NLO)~\cite{ref:MCLO}. 
Measurements of W + jet cross sections are an important test of QCD and
may be used to validate these new approaches. A good understanding of W + jet
production is vital to reduce the uncertainty on the background to top pair production
and to increase the sensitivity to higgs and new physics searches at the Tevatron
and the LHC.

This contribution describes a new measurement of the W + jet cross section as a function of 
relevant jet kinematic variables. 
%
Cross sections have been corrected
to particle level jets, and are defined within a limited W
decay phase space, closely matching that which is experimentally accessible. 
This definition, easily reproduced
theoretically, minimizes the model dependence that can
enter a correction back to the full W cross-section.
This analysis is based on $320 \pm 18$~${\rm pb^{-1}}$ of data 
collected by the CDF II detector at Tevatron collider.

The CDF II detector~\cite{ref:CDF} is an azimuthally and forward-backward symmetric
apparatus situated around the $p\bar{p}$ interaction region, consisting of a magnetic
spectrometer surrounded by calorimeters and muon chambers. 
\wev candidate events are selected from a high $E_T$ electron trigger ($E^e_T\ge18$ GeV, $|\eta^e|<1.1$) 
by requiring one good quality electron candidate ($E^e_T\ge20$ GeV) and the missing transverse energy 
($E\!\!\!/_T$) to be greater than $30$~${\rm GeV}$. 
%
%
%
The \wev candidate events are then classified according to their jet multiplicity into
four n$-$jet samples ($n\ge1,~4$). 
Jet are searched for using an iterative seed-based cone algorithm~\cite{ref:jetalg}, 
with a cone radius $R=\sqrt{(\Delta\eta)^2+(\Delta\phi)^2}=0.4$.
Jets are requested to have a corrected transverse energy $E_T^{jet}>15$GeV and a pseudorapidity
$|\eta|<2.0$. 
$E_T^{jet}$ is corrected on average for the
calorimeter response and the average contribution to the jet energy from additional $p\bar{p}$
interaction in the same bunch crossing~\cite{ref:jetnim}. 

Backgrounds
can be classified in two categories: QCD and W-like events. 
The latter is represented by events which manifest themselves as real 
electrons and/or $E\!\!\!/_T$ in the final state, namely: $W\to\tau\nu$, $Z\to e^+e^-$,
WW, top pair production. 
The former is mainly coming from jets production. 
While the W-like backgrounds are modeled with Monte Carlo
 simulations, the QCD background is described with a data-driven technique.
To extract the background fraction in each $W + \geq n{\rm -jet}$ sample
the  $E\!\!\!/_T$ distribution of candidates is fitted to background
and signal templates. 
\texttt{Alpgen}~\cite{ref:alpgen} interfaced to \texttt{HERWIG}~\cite{ref:HERWIG} has been used to generate
the $W\to\tau\nu$, $Z\to e^+e^-$ backgrounds and the W signal, \texttt{PYTHIA}~\cite{ref:PYTHIA}
have been used for top and WW backgrounds.
The sensitivity of these template
on the particular set of parton level cuts and Monte Carlo parameters has been
studied. It is always below a 5\% level and this effect has been included in the systematic
on the background estimate. The template for the QCD background 
is extracted from the data selecting a background enriched sample using candidate electron
satisfying all standard quality requirements but at least failing two of them.
Cross-checks of this method have been performed by looking to other W kinematic distributions 
as the transverse mass of the W $m_T^W$ and the electron $E_T^{e}$. In all these variables a very good agreement between data
and background models has been found.
 The total background fraction ranges from 1\% at low jet multiplicity and low $E_T^{jet}$ to
 80\% at high $E_T^{jet}$ and is largely dominated by the contribution of 
 QCD. At high jet multiplicity and high $E_T^{jet}$,
the contribution to the background from top production is sizeable ($\ge50$\%). In this region  
 the uncertainty on the top pair production cross section dominates the background systematic. 
 Elsewhere the main contribution to the uncertainty on the background fraction comes 
 from the limited statistic of the QCD background sample.

%
A full detector simulation has been used to take into account selection efficiencies, 
coming from geometric acceptance, electron identification and $E\!\!\!/_T$ and $E_T^{e}$ resolution effects.
%
%
The full CDF II detector simulation accurately reproduces electron acceptance and
identification inefficiencies: no evidence of a difference between data and
simulation have been found in the $Z\to e^{+}e^{-}$ sample.
 To minimize the theoretical uncertainty in the extrapolation of the measurement,
 the cross section has been defined for the W phase space 
 accessible by the CDF II detector: $E_T^{e}>20$GeV, $|\eta^{e}|<1.1$,  
 $E\!\!\!/_T>30$GeV and $m_T^{W}>20$GeV/c$^2$. 
This eliminates the dependence on Monte Carlo models to extrapolate 
the visible cross section to the full W phase space.
 Nevertheless Monte Carlo events have been used to correct for inefficiency and boundary
 effects on the kinematic selection that defines the cross section.
 Different Monte Carlo prescriptions
 have been checked and the critical parameters have been
 largely scanned. These effects turned out to be at the 5\% level at low $E_T^{jet}$.
 They have been included into the systematic uncertainty 
 on the efficiency which is $(60\pm3)$\%, largely independent of the jet kitematic. 

%
%
The candidate event yields, background fractions and efficiency factors are combined
to form the raw W + jet cross sections. The raw cross sections are then corrected 
back to the hadron level jet cross sections using Monte Carlo event samples. 
\texttt{Alpgen} interfaced with \texttt{PYTHIA-TUNE A}~\cite{ref:TUNEA} provides a 
reasonable description of the jet and underlying event properties, and is used to 
determine the correction factors, defined as the ratio of the hadron level cross section 
to the raw reconstructed cross section.

\begin{figure}
\includegraphics[height=.215\textheight]{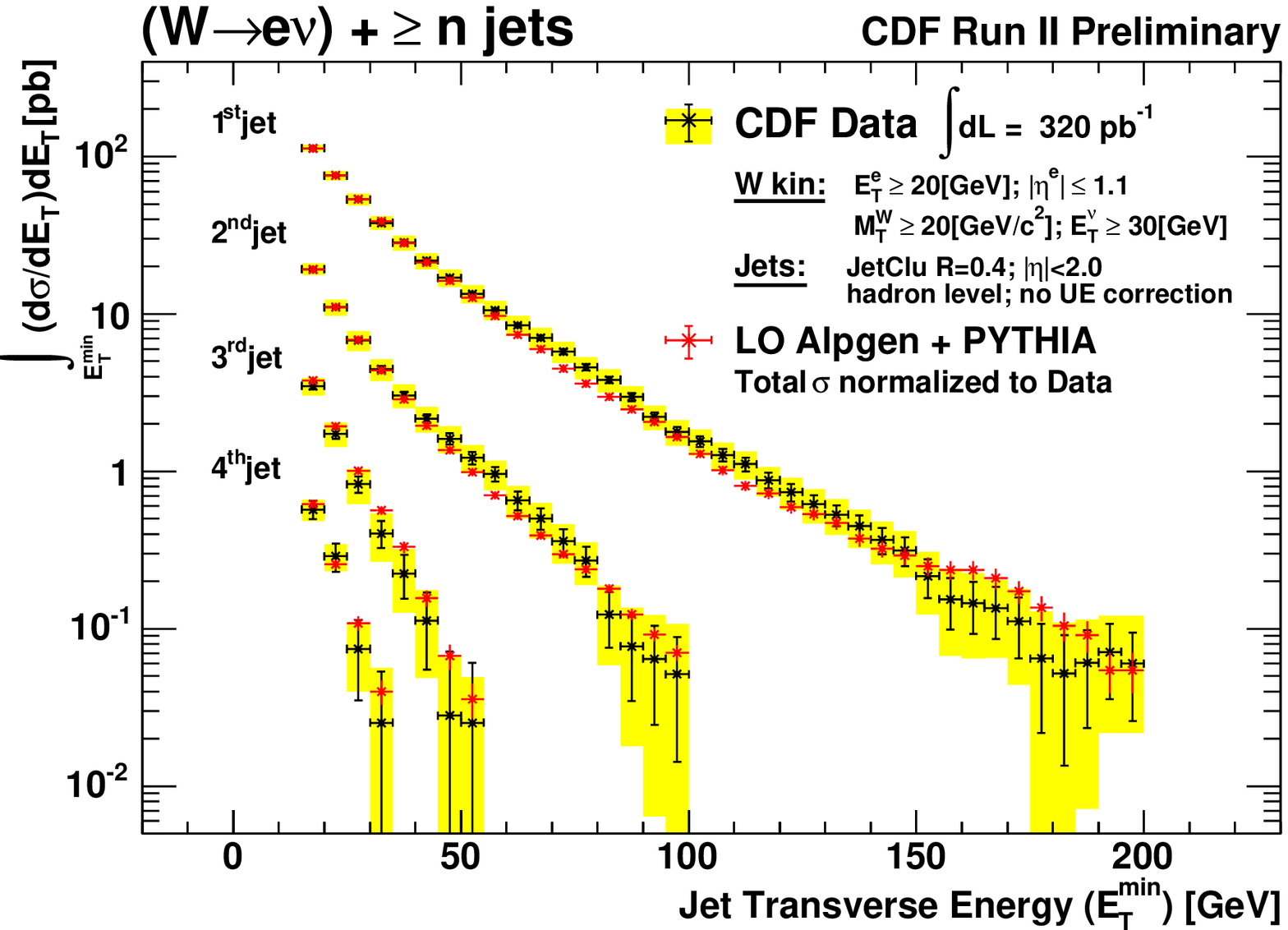}
\includegraphics[height=.215\textheight]{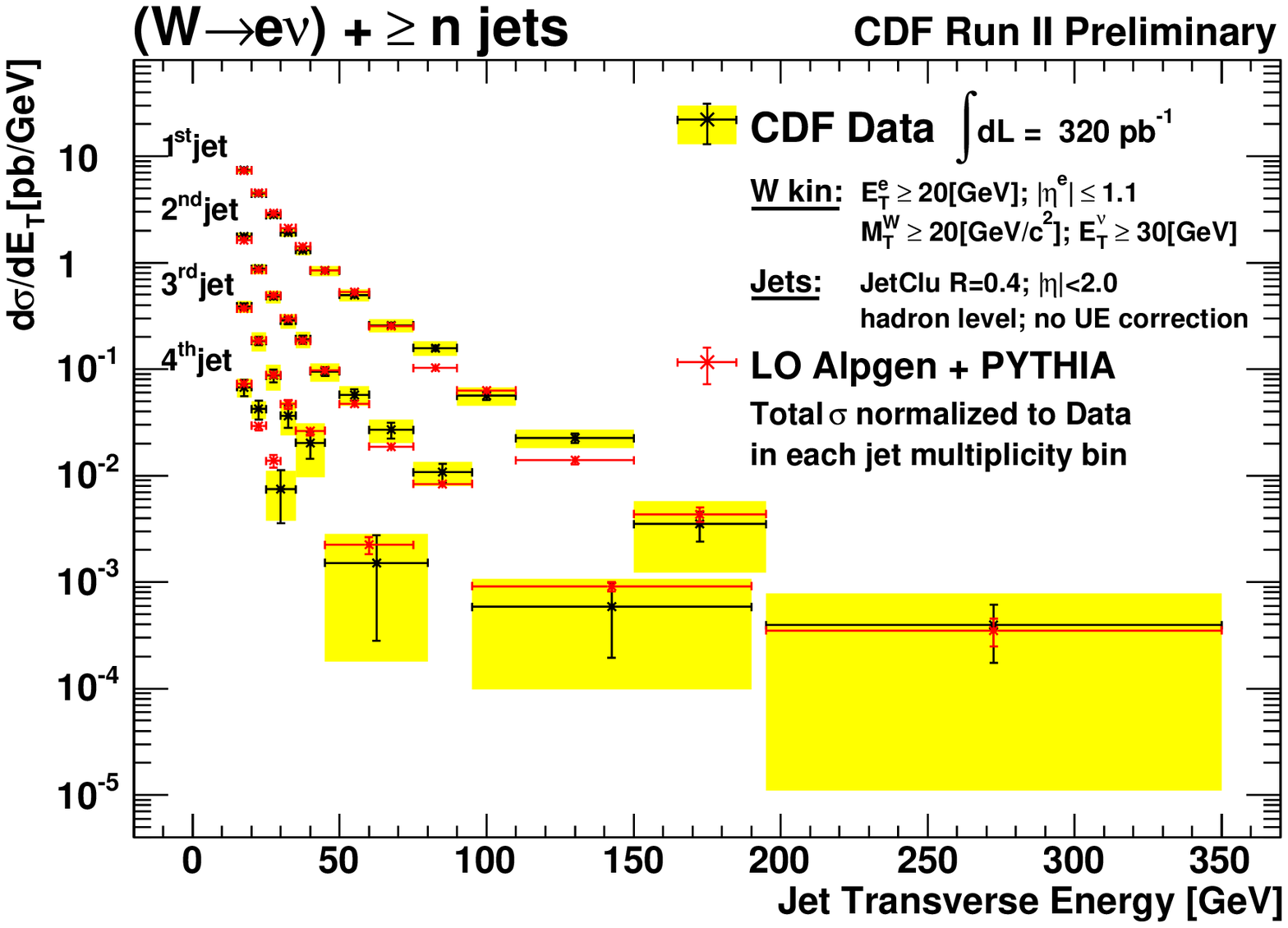}
\includegraphics[height=.215\textheight]{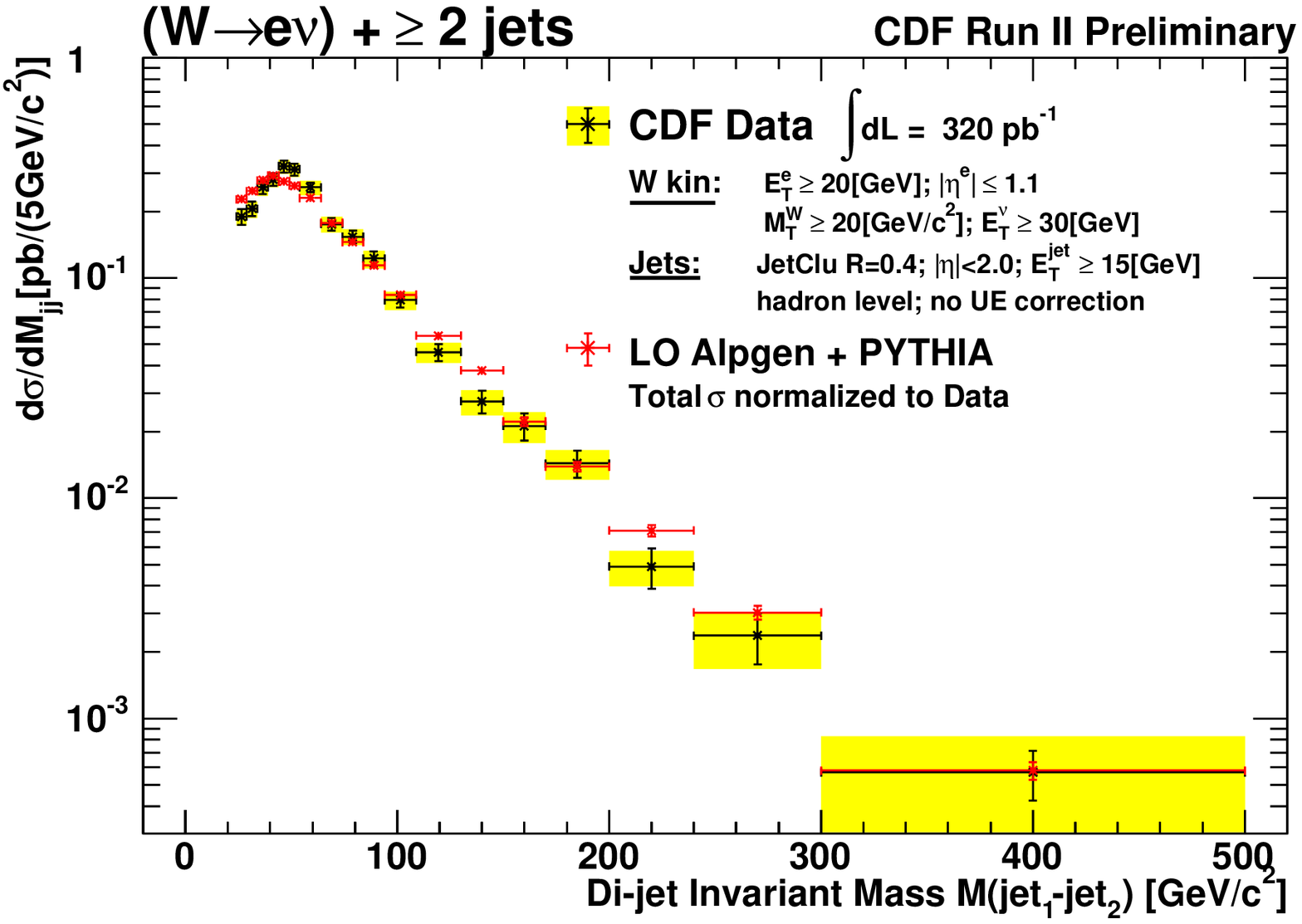}
\includegraphics[height=.218\textheight]{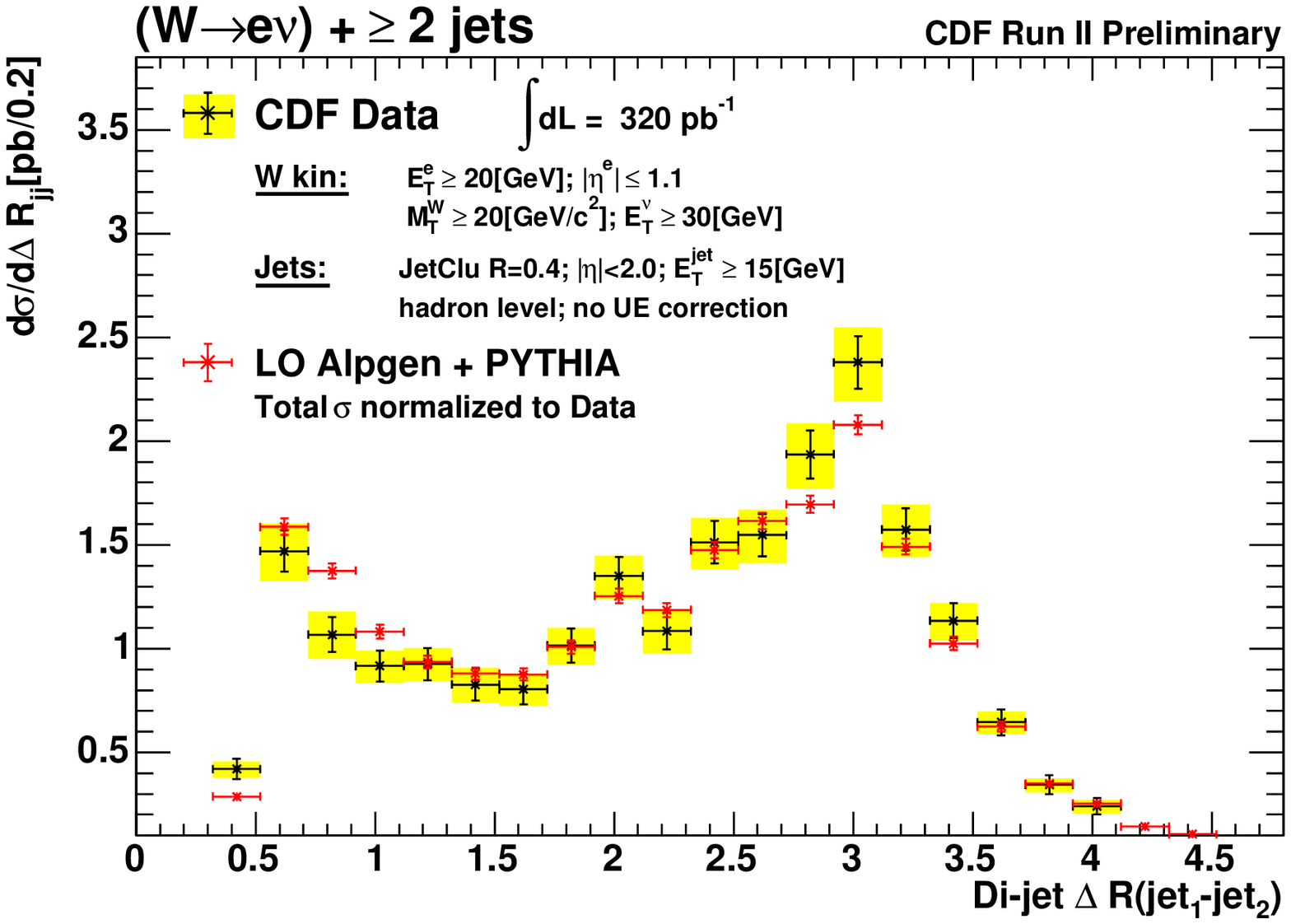}
\caption{Top: Cumulative cross section $\sigma(W\to e\nu+\geq n-{\rm jets}; E_{T}^{jet}(n) > E_{T}^{jet}(min))$
as a function of the minimum $E_T^{jet}(min)$ (Left) and differential cross section 
$d\sigma(W\to e\nu+\geq n-{\rm jets})/d E_{T}^{jet}$ (Right) for the first, second, third and fourth inclusive 
jet sample. Bottom: Differential cross section  $d\sigma(W\to e\nu+\geq 2-{\rm jets})/d M_{j1j2}$ (Left) and
$d\sigma(W\to e\nu+\geq 2-{\rm jets})/d R_{j1j2}$ (Right) respectively as a function of the invariant
mass and angular separation of the leading 2 jets. 
Data are compared to \texttt{Alpgen}+\texttt{PYTHIA} predictions normalized to
the measured cross section in each jet multiplicity sample.
\label{fig:xs}}
\end{figure}

The measured cross section are shown in fig.~\ref{fig:xs}. Results are presented as both cumulative
$\sigma(W\to e\nu+\geq n-{\rm jets}; E_{T}^{jet}(n) > E_{T}^{jet}(min))$
and differential
$d\sigma(W\to e\nu+\geq n-{\rm jets})/d E_{T}^{jet}$
distribution where $E_{T}^{jet}$ is that of the $i^{th}-$jet (Top plots fig.~\ref{fig:xs}).
The measurement spans over three orders of magnitude in cross section and close to 
$200$~${\rm GeV}$ in jet $E_{T}$ for the $\geq 1-{\rm jet}$ sample. For each jet multiplicity, 
the jet spectrum is reasonably well described by
individually normalized \texttt{Alpgen}+\texttt{PYTHIA} $W+n-{\rm parton}$ samples.
The shape of the dijet invariant mass and angular correlation (Bottom plots fig.~\ref{fig:xs}) are also
well modeled by the same theory prediction.
The systematic error is dominated by the uncertainty on the jet energy scale ($\sim 3\%$) at low $E_T^{jet}$,
while at high energy the dominant contribution comes from the uncertainty on the
background fraction, in particular from the limited statistic of the QCD background sample. 
We expect to reduce drastically this effect by increasing the statistic of the data sample. We are currently
working on similar measurements  in the $Z+\geq n-$jets events. This event sample, thanks to the low
background contamination and to the closed kinematic, is also particularly 
suitable to study the underlying event and the jet shape.


%


\end{document}